# Higgs Amplitude Mode in BCS Superconductors $Nb_{1-x}Ti_xN$ induced by Terahertz Pulse Excitation

Ryusuke Matsunaga[1], Yuki I. Hamada[1], Kazumasa Makise[2], Yoshinori Uzawa[3], Hirotaka Terai[2], Zhen Wang[2], and Ryo Shimano[1]

[1]*Department of Physics, The University of Tokyo, Tokyo, 113-0033, Japan*

[2]*National Institute of Information and Communications Technology, 588-2 Iwaoka, Nishi-ku, Kobe 651-2492, Japan*

[3]*National Astronomical Observatory of Japan, 2-21-1 Osawa, Mitaka, Tokyo 181-8588, Japan*

## Abstract

**Ultrafast responses of BCS superconductor $Nb_{1-x}Ti_xN$ films in a nonadiabatic excitation regime were investigated by using terahertz (THz) pump-THz probe spectroscopy. After an instantaneous excitation with the monocycle THz pump pulse, a transient oscillation emerges in the electromagnetic response in the BCS gap energy region. The oscillation frequency coincides with the asymptotic value of the BCS gap energy, indicating the appearance of the theoretically-anticipated collective amplitude mode of the order parameter, namely the Higgs amplitude mode. Our result opens a new pathway to the ultrafast manipulation of the superconducting order parameter by optical means.**

With spontaneous breaking of continuous symmetry, two types of collective excitations associated with the order parameter emerge. One is the gapless phase mode called as the Nambu-Goldstone mode, and the other is the gapped amplitude mode also referred to as the Higgs mode from the analogy to the Higgs boson in particle physics [1, 2], as schematically shown in Fig. 1(a). Recently, the Higgs amplitude mode has been observed in strongly interacting superfluid phases of bosonic ultracold atoms in optical lattices by means of Bragg spectroscopy [3] and lattice modulation [4]. The studies of the Higgs mode realized on table top experiments would provide substantial platforms for exploring the nature of symmetry-broken states in quantum many-body physics. In condensed matter systems, the amplitude mode has been widely observed in charged density wave (CDW) systems by Raman or pump-probe spectroscopy [5-8] and in an antiferromagnet by neutron spectroscopy [9]. However, the observation of the amplitude mode in fermionic condensates has been limited to the specific cases of superconducting CDW compound $NbSe_2$ [10, 11] and $p$-wave superfluid $^3He$ [12, 13]. Then, we can pose a question as to whether the Higgs mode in a pure metallic BCS superconductor (SC), which does not couple to the radiation field, can be observed experimentally.

The amplitude mode in the BCS order parameter has been anticipated to appear in a response to a fast perturbation in nonadiabatic regime [14-23]. Depending on the perturbation strength, the nonequilibrium dynamics would exhibit a persistent oscillation, a transient oscillation obeying a power-law decay, or a quantum quench of the order parameter which cannot be described by the time-dependent Ginzburg-Landau theory or the Boltzmann equation [16, 17]. A sudden switching of the pairing interaction by using Feshbach resonance in ultracold atoms [24] is one promising way to realize such a nonequilibrium state, while it still remains experimentally challenging. An alternative way to induce the transient oscillation of the order parameter has been proposed in conventional metallic BCS SCs [19]. When a BCS ground state is nonadiabatically excited by a short laser pulse, the coherence between different quasiparticle (QP) states leads to the oscillation of the order parameter. Such a nonadiabatic excitation for BCS superconductivity requires a short pump pulse with the duration $\tau_{pump}$ small enough compared to the response time of the BCS state characterized by the BCS gap $\Delta$ as $\tau_\Delta=\hbar/\Delta^{-1}$. Here a near-visible femtosecond optical pulse is not applicable, because the huge excess energies of photoexcited hot electrons in the order of electronvolts are transferred to the generation of large amounts of high-frequency phonons ($\hbar\omega>2\Delta$), which in turn induce the Cooper pair breaking. This process destroys the nonadiabatic excitation condition even if one uses the laser pulse much shorter than $\tau_\Delta$ [25, 26]. Therefore, to ensure the nonadiabatic excitation, it is necessary to use a short pump pulse with its photon energy resonant to the BCS gap which is typically located in terahertz (THz) frequency range [19]. With the recent development of the THz technology, such an intense and monocycle-like THz pulse has become available [27], making it possible to investigate the THz nonlinear response in a variety of materials [28-32]. In a *s*-wave SC of NbN film, the ultrafast pair-breaking and the following QP dynamics have been investigated by the intense THz pump-THz probe (TPTP) spectroscopy [26]. Nonlinear THz transmission experiments in NbN have also been reported recently [33, 34].

In this Letter, we investigated the coherent transient dynamics of superconducting $Nb_{1-x}Ti_xN$ films after the THz pulse excitation in the nonadiabatic excitation regime. The time-domain oscillation of the order parameter was observed in the pump-probe delay dependence of the transmitted probe THz electric field (*E*-field). The oscillation frequency is in excellent agreement with the theoretical predictions.

The output from a regenerative amplified Ti:sapphire laser system with 800-nm center wavelength, 1-mJ pulse energy, 90-fs pulse duration, and 1-kHz repetition rate was divided into three beams; for the generation of the pump and probe THz pulses and for the gate pulses for the electro-optic (EO) sampling of the transmitted probe THz pulse. The intense pump THz pulse was generated by the tilted-pulse-front method with a $LiNbO_3$ crystal [27], and the detail of our experimental configuration was described in Ref. [35]. The pump pulse width defined by FWHM of the envelope curve of the *E*-field amplitude was $\tau_{pump}\sim1.5$ ps. The probe THz pulse was generated by the optical rectification in a ZnTe crystal. As schematically shown in Fig. 1(b), a wire grid polarizer (WGP) inserted in the optical path of the pump THz pulse ($E_{pump}//x$) reflects the probe THz pulse ($E_{probe}//y$) so that the pump and probe THz pulses are collinearly irradiated to the sample. Another WGP was placed after the sample to block the pump THz pulse and to transmit the probe THz pulse only. The waveform of the probe $E$-field was detected by the EO sampling in a ZnTe crystal. By scanning both the delay time of the gate pulse to the probe THz pulse, $t_{\text{gate}}$, and the delay time of the probe to the pump THz pulse, $t_{\text{pp}}$, we recorded the probe THz $E$-field $E_{\text{probe}}(t_{\text{gate}}, t_{\text{pp}})$ in the two-dimensional time domains of $t_{\text{gate}}$ and $t_{\text{pp}}$ [36]. The details in our two-dimensional THz time-domain spectroscopy system were described in the previous paper [26].

The $Nb_{1-x}Ti_xN$ films were fabricated on fused quartz (FQ) or MgO substrates using the dc reactive sputtering method [37]. We used three different samples; (sample A) $x$=0.2 and film thickness $d$=12 nm on a 1 mm-thick FQ, (sample B) $x$=0.2 and $d$=30 nm on a 0.5 mm-thick FQ, and (sample C) $x$=0 and $d$=24 nm on a 0.5 mm-thick MgO. Figure 1(c) shows the temperature dependence of the real-part optical conductivity spectra $\sigma_1(\omega)$ of the sample C without the THz pump. The solid curves are calculated by the Mattis-Bardeen model with arbitrary electron mean-free path [38, 39] to evaluate the gap energy at each temperature. The temperature dependence of the gap energy is shown in Fig. 1(d). The BCS gap energies at 4 K are evaluated as $2\Delta_0$=0.72, 1.1, and 1.3 THz, for the samples A, B, and C, respectively, which gives the ratio $\tau_{\text{pump}}/\tau_\Delta$ as 0.57(A), 0.81(B), and 0.98(C).

Figure 1(e) shows the time-domain waveform of the probe THz pulse, $E_{\text{probe}}(t_{\text{gate}})$, transmitted after the sample A below $T_c$=8.5 K without the THz pump. As indicated by the vertical line in Fig. 1(e), the probe $E$-field at $t_{\text{gate}}$=2.1 ps ($\equiv t_0$) sensitively indicates the growth of the superconducting state. In fact, as shown by Fig. 1(f), the value $E_{\text{probe}}$ at $t_{\text{gate}}$=$t_0$ shows one-to-one correspondence with the BCS gap energy $2\Delta$ obtained from Fig. 1(d). Therefore, in order to detect the temporal evolution of the order parameter $\Delta(t_{\text{pp}})$ after the pump, we monitored the probe $E$-field at this fixed gate delay time, $E_{\text{probe}}(t_{\text{gate}}=t_0, t_{\text{pp}})$. Note that, this correspondence between the gap energy $2\Delta$ and $E_{\text{probe}}(t_{\text{gate}}=t_0, t_{\text{pp}})$ in the equilibrium condition without the pump does not necessarily hold in the nonequilibrium case. Therefore, we numerically confirmed that $E_{\text{probe}}(t_{\text{gate}}=t_0, t_{\text{pp}})$ indeed reflects the transient behavior of the order parameter changing in a time scale of $\tau_\Delta$. The details are given in Supplemental Material [40].

Figure 2(a) shows the temporal evolution of the change of the probe $E$-field, $\delta E_{\text{probe}}$, at $t_{\text{gate}}=t_0$ as a function of $t_{\text{pp}}$ in the sample A with $\tau_{\text{pump}}/\tau_\Delta=0.57$ for various pump intensities. After a fast rise within 2 ps which we will discuss later, an oscillatory behavior is clearly identified. As the pump intensity increases, the oscillation amplitude increases and the frequency decreases, and the oscillation is heavily damped in the strong excitation limit. At each excitation level, $\delta E_{\text{probe}}$ asymptotically reaches to a constant value accompanied by the damped oscillation. Besides the oscillation, $\delta E_{\text{probe}}$ shows a slow increase at $t_{\text{pp}}>2$ ps to the constant value, indicating the gradual decrease of the gap energy. Such a slow decrease of the gap energy after the pump pulse irradiation has also been observed in the previous near-visible optical pump experiments, where the excess photon energy of the pump pulse gives rise to the generation of phonons which in turn causes the pair breaking in a slower time scale [25, 26]. Meanwhile, a recent calculation using the nonequilibrium dynamical mean-field theory [23] has also showed that such a slow thermalization dynamics can occur as a unique character of a nonequilibrium state, even without taking into account the interaction with the phonon system. In the present experiment, whereas the central photon energy of the pump THz pulse is resonant to the gap energy, the high-frequency components of the pump THz pulse larger than the gap energy bring the excess energy to the QP system. Therefore the slow increase in Fig. 2(a) can be attributed to the thermalization process of the excess energy.

As shown by the solid curves in Fig. 2(a), the oscillating part of $\delta E_{\text{probe}}(t_{\text{pp}})$ is fitted by the following equation

$$\delta E_{\text{probe}}(t_{\text{pp}}) = C_1 + C_2 t_{\text{pp}} + a \frac{\cos(2\pi f t_{\text{pp}} + \varphi)}{(t_{\text{pp}} - t')^b}, \qquad (1)$$

where $C_1$, $C_2$, $a$, $b$, $\varphi$, $f$, and $t'$ are parameters. The first term indicates the non-oscillating part of the gap energy. The second term is introduced to reproduce the gradual decrease of the gap energy, which is attributed to the thermalization process as described above. The third term describes the order parameter oscillation with the power-law decay as theoretically predicted [14, 16, 17]. Figure 2(b) shows the oscillation frequency $f$ obtained from the fits at various pump intensities. Here we also plot the values of $2\Delta$ at $t_{pp}$=8 ps where the oscillation is damped, which indicates the asymptotic value $2\Delta_\infty$ of the gap energy after the pump. Because of the slow change of the order parameter in this temporal region, we evaluated $2\Delta_\infty$ from the observed $\delta E_{probe}(t_{pp}$=8 ps) by using the correspondence in Fig. 1(f). The decrease of $2\Delta_\infty$ as a function of the pump intensity represented in Fig. 2(b) is reasonable because the increase of the excited QP density causes the gap reduction. The fitted values $f$ and their pump-intensity dependence are in excellent agreement with $2\Delta_\infty$, which is a characteristic feature of the order parameter oscillation predicted in the theoretical studies [16, 17]. Therefore, this result strongly suggests that the oscillatory signal arises from the collective Higgs amplitude mode anticipated in the nonadiabatic excitation condition. Note that the oscillatory signal is observed in the cross-linear polarization configuration of the TPTP experiments, which also indicates its origin as the Higgs mode of isotropic $s$-wave SCs. It is intriguing that the polarization dependent TPTP experiments would elucidate the nature of symmetry of such collective modes.

Figure 2(c) shows the fitted parameter $b$, the power-law index for decay of the oscillation, as a function of the pump intensity. The theoretical studies have shown that within the linear approximation the oscillation decays with $b$=0.5 for the weak-coupling BCS case due to the mixing of the collective mode and QP states [14-16], and with $b$=1.5 for the strong-coupling case [21]. Our result shows that $b$ changes from about 1 to 3 depending on the pump intensity. Such a rapid decay depending on the excitation intensity could be considered as a signature of the overdamped oscillation of the order parameter [16, 17].

The dynamics after the THz pulse excitation was also investigated in the frequency domain. Figure 3(a) shows the temporal evolution of the real-part optical conductivity spectra $\sigma_1(\omega)$ as a function of $t_{pp}$, obtained from the TPTP spectroscopy in the two-dimensional time domains. The optical conductivity spectrum $\sigma_1(\omega; t_{pp})$ at each delay time $t_{pp}$ was calculated from the waveform of the transmitted probe $E$-field. Figure 3(b) shows the $\sigma_1(\omega)$ spectra at each $t_{pp}$ indicated by the white dotted lines in Fig. 3(a). For comparison, Fig. 3(b) also shows the $\sigma_1(\omega)$ spectra before the pump ($t_{pp}$ =-2 ps) as the black dotted curves. The temporal oscillation of the conductivity spectrum is clearly seen, suggesting the oscillation of the gap energy. However, the oscillation of the onset of the gap is not clear, which might be obscured by the smooth onset of the conductivity gap as observed even without the pump in our film samples. On the other hand, the spectral weight clearly shows the temporal oscillation and so as the superfluid density. Therefore we consider that the results can be interpreted as the order parameter oscillation.

We also performed the TPTP experiments in the samples B and C that exhibit larger BCS gap energies than the sample A. Figures 4(a) and 4(b) show the temporal evolution of $\delta E_{\text{probe}}$ as a function of $t_{\text{pp}}$ in the samples B and C with $\tau_{\text{pump}}/\tau_\Delta$=0.81 and 0.98, respectively, for various pump intensities. The regions surrounded by the dotted lines in Figs. 4(a) and 4(b) are enlarged in the insets with the fitted curves. As similar to the case in Fig. 2(a), $\delta E_{\text{probe}}$ sharply increases and decays into a nearly constant value within 2 ps, on top of which the long-lived oscillation is identified at $t_{\text{pp}}$>2 ps. Therefore, the mechanism of this initial peak-like signal should be different from the order parameter oscillation, and we excluded this temporal domain in the analysis in Fig. 2(a). Whereas the microscopic description for this initial process still remains as an issue to be resolved, the pump THz pulse is considered to result in a direct generation of high density QPs within this time scale without mediating the phonon excitation [26]. Then the initial population of QPs should decrease due to the recombination process until the QP system reaches in a chemical equilibrium with the phonon subsystem. Aside from the oscillation, $\delta E_{\text{probe}}$ becomes almost constant at $t_{\text{pp}}$>2 ps, which suggests that the two systems are equilibrated within 2 ps due to the high QP recombination rate. Thus the rapid decay of $\delta E_{\text{probe}}$ within 2 ps may be attributed to the equilibration process with the phonon system. The appearance of the Higgs amplitude mode in the temporal region where the QPs equilibrate with the phonon subsystem suggests its nature of the collective mode, being robust against the much faster scattering processes of individual QPs with phonons.

The oscillation at $t_{\text{pp}}$>2 ps in Fig. 4 is hardly identified as the pump intensity increases. Our previous experiments in the samples whose gap energy is similar to sample C have been performed in such a high excitation regime [26], where the spatial inhomogeneity appears. The fitted oscillation frequencies $f$ in the weak excitation limit (insets of Figs. 4(a) and 4(b)) are higher than that in Fig. 2(a), corresponding to the large gap energies in the sample B and C. We also found that the gradual increase of $\delta E_{\text{probe}}$ at $t_{\text{pp}}>2$ ps seen in the sample A (Fig. 2(a)) was not observed in the sample C. This is ascribed to the small excess photon energy of the pump pulse in the sample C with the large BCS gap energy. Compared with the result in sample A (Fig. 2(a)), the oscillation becomes less prominent for the samples B and C. This result is attributed to two reasons. First, the larger values of $\tau_{\text{pump}}/\tau_\Delta$ indicate that the nonadiabaticity is weak for the sample B and C. In addition, the oscillation is considered to be smeared because the oscillation cycle ($1/f\sim 0.75$ ps in the sample C) is comparable to the temporal resolution in this measurement [40].

In conclusion, we investigated the ultrafast dynamics of $s$-wave SCs in a nonadiabatic excitation regime by using the monocycle THz pulse. A clear temporal oscillation was observed in the transmission of the probe THz pulse, whose oscillation frequency is in excellent accordance with the value of the asymptotic gap energy in the quasi-stable nonequilibrium state. The results are well accounted for by the theoretically-anticipated BCS order parameter oscillation, namely the collective Higgs amplitude mode. Since the collective amplitude mode can be thought of as the collective Rabi oscillation [15], the phenomenon may serve for quantum optical applications in the THz regime. The extension of the measurement to other types of superconductivity such as high-$T_{\text{c}}$ SCs with anisotropic symmetry [41] is also highly intriguing. Our result would pave a way for ultrafast optical manipulation of order parameter in SCs.

We acknowledge N. Tsuji and H. Aoki for fruitful discussions. This work was partially supported by a Grant-in-Aid for Scientific Research (Grants Nos. 25800175, 22244036 and 20110005) and by the Photon Frontier Network Program from MEXT, Japan.

**Figure captions**

**Fig. 1** (Color online) (a) A schematic picture of the phase mode (blue arrow) and the amplitude mode (red arrow) represented by the effective potential in the plane of complex order parameter $\Psi$. (b) Schematic configuration of the TPTP spectroscopy. WGP: a wire grid polarizer. (c) Temperature dependence of the real-part optical conductivity spectra in the sample C without the pump. The solid curves are calculated by the Mattis-Bardeen model. (d) Temperature dependences of the BCS gap energies for the samples A, B, and C. (e) The waveforms of the probe THz $E$-field $E_{probe}$ as a function of the gate delay time $t_{gate}$ at various temperatures without the pump. (f) The temperature dependence of the BCS gap $2\Delta$ in equilibrium and $E_{probe}$ at the fixed delay time of $t_{gate}=2.1$ ps ($=t_0$) indicated by the vertical line in (e) for the sample A.

**Fig. 2** (Color online) (a) The open circles show the temporal evolution of the change of the probe $E$-field, $\delta E_{probe}$, at $t_{gate}=t_0$ as a function of $t_{pp}$ in the sample A at 4 K. The solid curves show the fitted results with Eq. (1). (b) The oscillation frequency $f$ obtained from the fits and the asymptotic gap energy $2\Delta_\infty$ as a function of the pump intensity. (c) The power-law decay index $b$ as a function of the pump intensity.

**Fig. 3** (Color online) (a) The temporal evolution of the real-part optical conductivity spectra $\sigma_1(\omega)$ with the peak $E$-field of 10 kV/cm of the pump THz pulse at 4 K as functions of the frequency and $t_{pp}$. (b) $\sigma_1(\omega)$ spectra at each $t_{pp}$ indicated by the dotted lines in (a). The black curves show the spectrum before the pump ($t_{pp}=-2$ ps). The arrows are guide for the eyes to the oscillation.

**Fig. 4** (Color online) (a) and (b) The temporal evolution of $\delta E_{probe}$ at a fixed delay $t_{gate}=t_0$ as a function of $t_{pp}$ in the sample B and C at 4 K, respectively, for various pump intensities. The regions surrounded by the dotted lines are enlarged in the insets with the fitted curves and the oscillation frequencies.

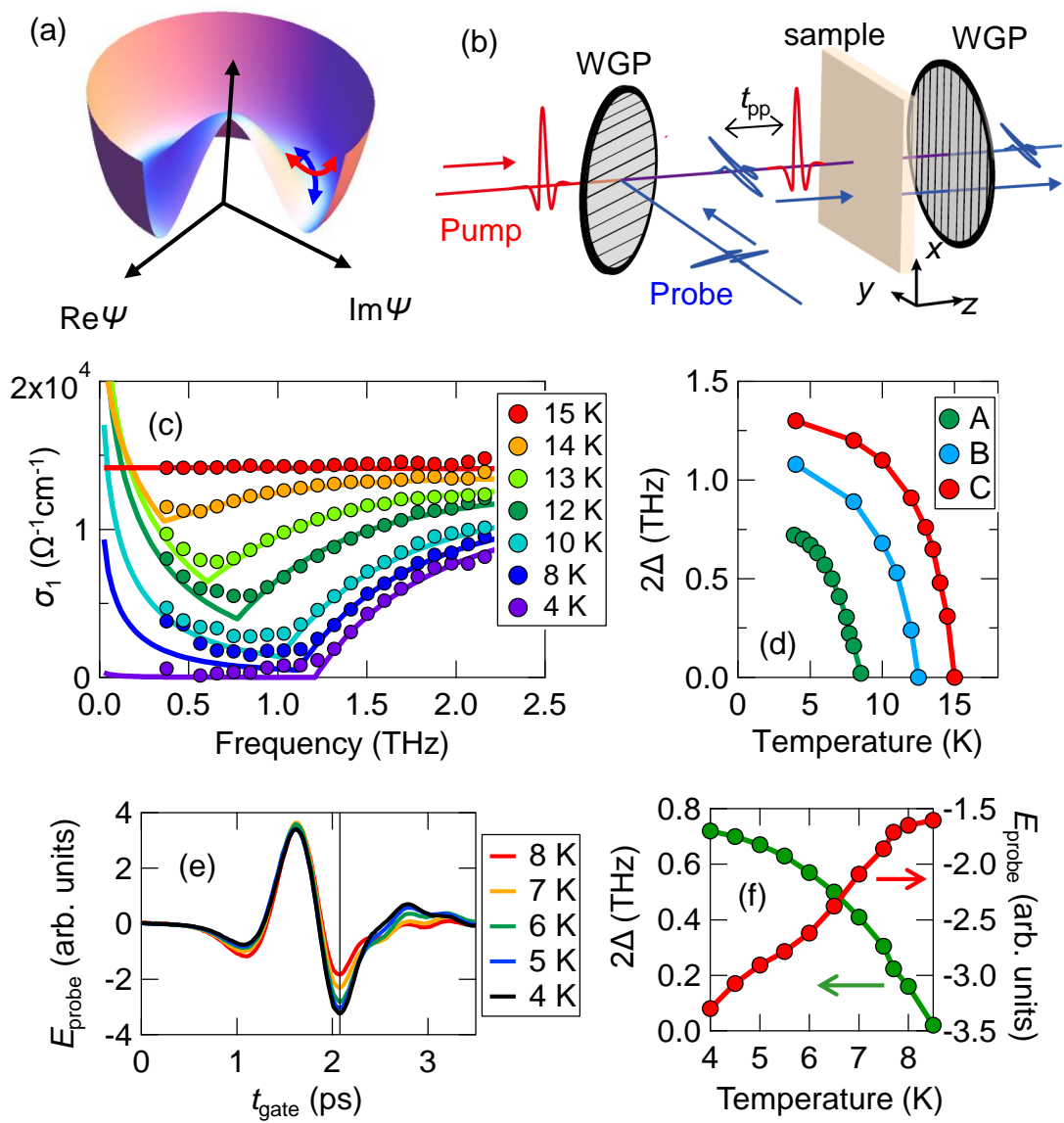

Figure 1

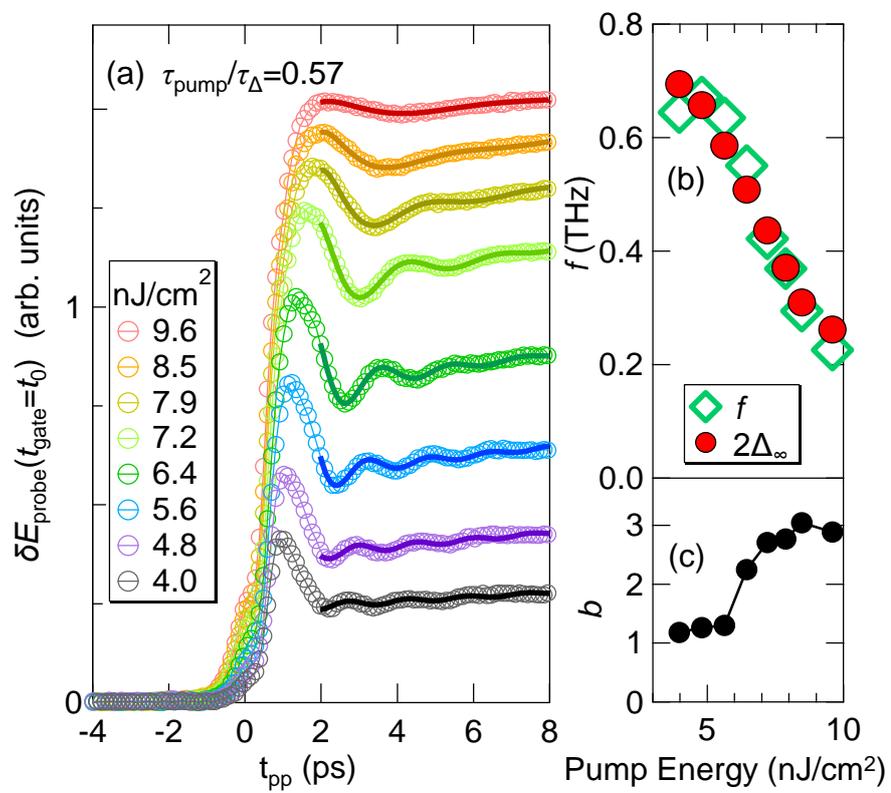

Figure 2

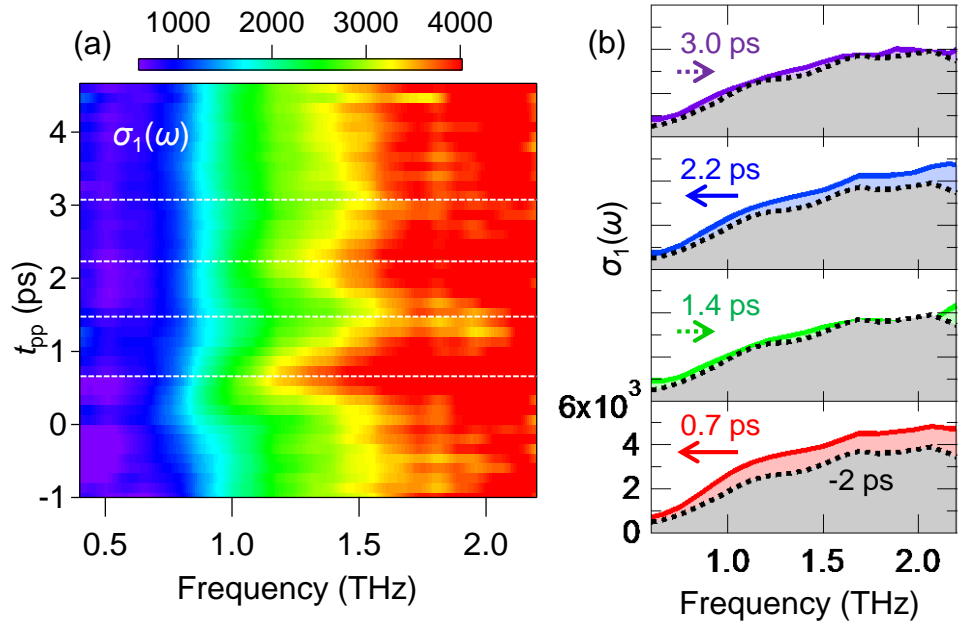

Figure 3

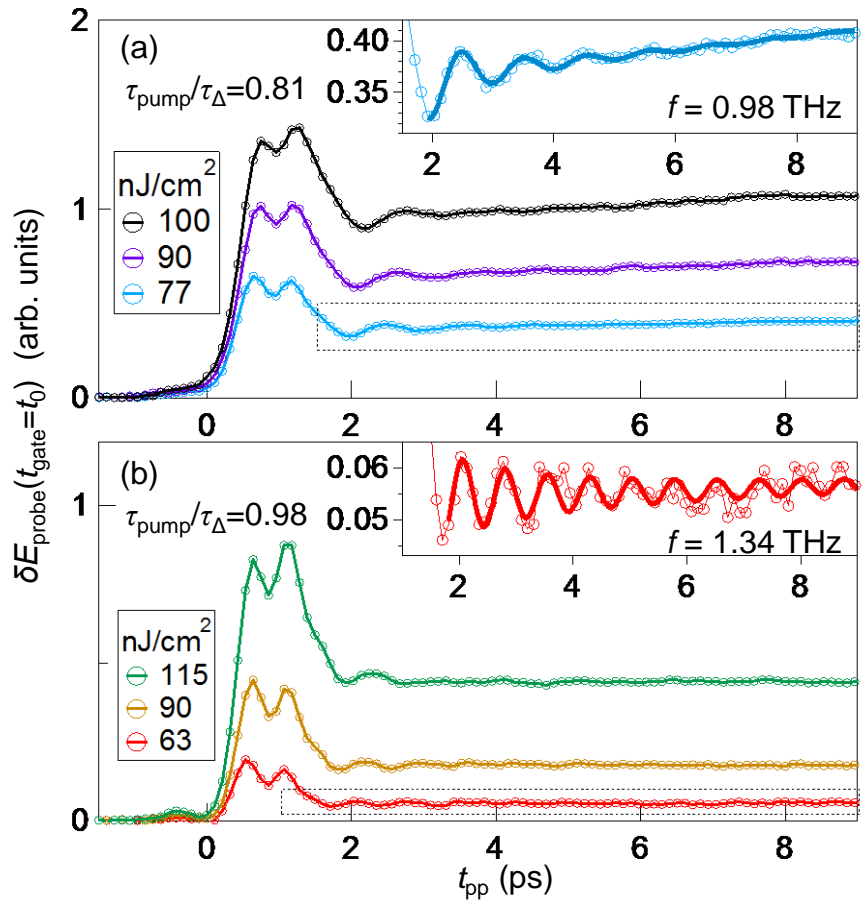

Figure 4